\documentclass[prl,aps,twocolumn,groupedaddress,floats,showpacs,final,superscriptaddress]{revtex4-1}
\usepackage[latin3]{inputenc}
\usepackage[makeroom]{cancel}
\usepackage{graphicx}
\usepackage{amsmath}
\usepackage{amsfonts}
\usepackage{amssymb}
\usepackage{chngcntr}
\usepackage{color}
\usepackage[left=2cm,right=2cm,top=2cm,bottom=2cm]{geometry}

\usepackage{graphicx} 
\usepackage{dcolumn}
\usepackage{bm}
\usepackage{simplewick}
\usepackage{array}
\usepackage{appendix}

\RequirePackage[
   hyperindex,colorlinks,bookmarksnumbered,
   plainpages=true,pdfstartview=FitH]{hyperref}
\hypersetup{linkcolor=blue,urlcolor=blue,citecolor=blue}
\usepackage{hyperref}

\definecolor{purple}{rgb}{0.5,0,0.6}

\usepackage{ulem}

\begin{document}

\date{\today}
\title{Nano-mechanics driven by Andreev tunneling}
\author{A. V. Parafilo}
\affiliation{Center for Theoretical Physics of Complex Systems, Institute for Basic Science, Expo-ro, 55, Yuseong-gu, Daejeon 34126, Republic of Korea}

\author{L. Y. Gorelik}
\affiliation{Department of 
Physics, Chalmers University of
Technology, SE-412 96 G{\" o}teborg, Sweden}

\author{M. Fistul}
\affiliation{Center for Theoretical Physics of Complex Systems, Institute for Basic Science, Expo-ro, 55, Yuseong-gu, Daejeon 34126, Republic of Korea}
\affiliation{Theoretische Physik III, Ruhr-University Bochum, Bochum 44801 Germany}
\affiliation{National University of Science and Technology "MISIS", Russian Quantum Center, Moscow 119049, Russia}

\author{H. C. Park}
\affiliation{Center for Theoretical Physics of Complex Systems, Institute for Basic Science, Expo-ro, 55, Yuseong-gu, Daejeon 34126, Republic of Korea}

\author{R. I. Shekhter}
\affiliation{Department of Physics, University of Gothenburg, SE-412
96 G{\" o}teborg, Sweden}

\date{\today}

\begin{abstract}
{ We predict and analyze mechanical instability and 
corresponding self-sustained mechanical oscillations occurring in a nanoelectromechanical system composed of a metallic carbon nanotube 
(CNT) suspended between two superconducting leads
and coupled to a scanning tunneling microscope (STM) tip.
We show that such phenomena are realized in the presence of both the 
coherent Andreev tunneling between the CNT and superconducting leads, 
and an incoherent single electron tunneling between the voltage biased 
STM tip and CNT. Treating the CNT as a single-level quantum dot, we 
demonstrate that the mechanical instability is controlled by the 
Josephson phase difference, relative position of the electron energy 
level, and the direction of the charge flow.
It is found numerically that  the emergence of the self-sustained 
oscillations leads to a substantial suppression of DC electric current.}
\end{abstract}
\maketitle 

{\it Introduction.} Modern nanomechanical resonators \cite{fundamentals} characterized by low damping and fine-tuning of the resonant frequency are widely used nowadays as supersensitive quantum detectors \cite{sensing}-\cite{huttelnano} and as the mechanical component for various nanoelectromechanical systems (NEMS) \cite{NEMS},\cite{blencowe}. The latter represent a promising platform for studying the fundamental phenomena generated by the quantum-mechanical interplay between nanomechanical resonator and electronic subsystem \cite{ekinci},\cite{cleland}.


Large amount of fascinating physical phenomena have been {predicted and observed} in various NEMS, e.g. energy level quantization of a nanomechanical oscillator \cite{quantlevelsSQ}, a {strong resonant coupling} of nanomechanical oscillator to  superconducting qubits \cite{laserCoolSQ}, mechanical cooling \cite{coolingexp,nqds,gorelikcooling}, a single-atom lasing effect \cite{atomlasing,laserCoolSQ}, {mechanical transportation of Cooper pairs \cite{shuttlenature}} and the generation of self-driven mechanical oscillations by a DC charge flow \cite{shuttle,blanter,huttel1,huttel2,huttel3,self}, just to name a few. 

{Significant part} of these effects are based on the resonant excitation of low damped mechanical modes by \textit{coherent } quantum dynamics occurring in the electronic subsystem. 
A straightforward method to establish coherent {quantum dynamics in mesoscopic devices}, e.g., the quantum beats, the microwave induced Rabi oscillations etc.,  is to use the macroscopic phase coherence of superconducting (SC) elements incorporated into NEMS, see, for example, the review \cite{parafiloreview}.
{In particular, in superconducting hybrid junctions \cite{flensberg}-\cite{baransky} the coherent electronic transport is determined by the presence of Andreev bound states \cite{andreevlevel},\cite{andreevlevel2}}. The applied DC or AC currents induce the  transitions between Andreev bound states, and the coherent high-frequency oscillations in an electronic subsystem occur \cite{gorelikcooling}. These coherent charge oscillations can excite the mechanical modes in the resonant limit only, {when} the frequency of mechanical mode matches Andreev energy level difference.

On other hand, an \textit{incoherent} quantum dynamics 
occurring in the electronic subsystem can induce the mechanical instability and subsequent formation of the self-driven mechanical oscillations in hybrid junctions.  
Incoherent quantum fluctuations of electric charge can be easily mediated {by tunneling of a single electron}.  The self-driven oscillations generated by a DC electronic flow have been predicted in \cite{shuttle,blanter}, later observed in a carbon nanotube (CNT) based transistor \cite{huttel1}, and studied in detail \cite{huttel2},\cite{huttel3}, {see, e.g., \cite{self} for recent experiment.} 


A nontrivial interplay between coherent and incoherent electric charge variation and its influence on the performance of NEMS can be achieved in a {\it nanomechanical Andreev device}, where normal and SC metals are bridged by a mechanically active mediator. 

In this Letter, we present a particular NEMS setup where the mechanical oscillations are strongly affected by a weak coupling to the electronic part of a system. We demonstrate that in the adiabatic limit as the frequency of mechanical oscillations is much smaller than the typical frequencies of electron dynamics, simultaneous presence of coherent Andreev tunneling and incoherent single electron tunneling {can induce mechanical instability of the resonator and result in the appearance of the self-sustained mechanical oscillations. }

\begin{figure}
\centering
\includegraphics[width=1.\columnwidth]{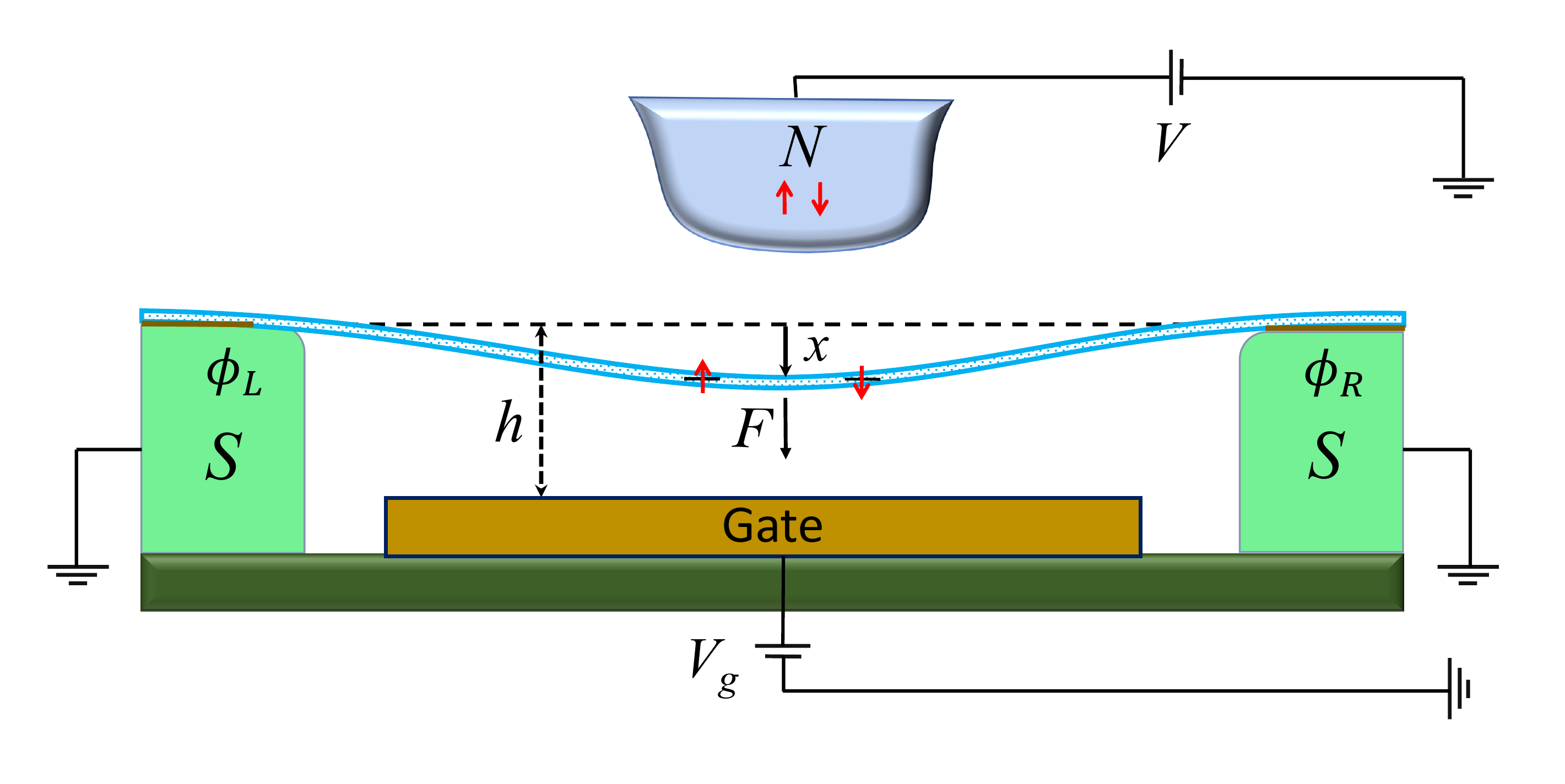}
\caption{Scheme of the superconducting (SC) nanoelectromechanical device. A single-wall carbon nanotube (CNT) is suspended between two SC leads which are characterized by the phases of SC order parameter, $\phi_{L,R}$. A normal metal electrode (STM tip) placed near the CNT-QD allows to inject electrons in CNT.  The nanoelectromechanical force $F$ between the CNT and gate electrode, which is located on the distance $h$ from the CNT, is controlled by a gate voltage $V_g$.
} 
\label{Fig1}
\end{figure}

{\it Model.}  We consider a metallic single-wall carbon nanotube suspended between two grounded SC electrodes and coupled to a scanning tunneling microscope (STM) tip via electron tunneling.
The two SC electrodes  are characterized by the same modulus $\Delta$ and different phases $\phi_{L,R}$ of SC order parameter,  and corresponding Josephson phase difference, $\phi=\phi_R-\phi_L$. We study the case where the CNT mean-level spacing is greater than temperature $k_BT$ and the bias-voltage $eV$ applied between STM tip and CNT. 
It allows us to treat the CNT as a movable single-level quantum dot (QD). The capacitive coupling between the CNT and a gate is controlled by a gate voltage $V_g$.
We aslo assume the dynamics of the CNT bending is reduced to the dynamics of the fundamental flexural mode.  The scheme of the described model is presented in Fig.\ref{Fig1}.

The Hamiltonian of the model reads as follows
\begin{eqnarray}\label{hamiltonian}
H=H_{N}+H_S+H_{CNT}+H_{tun}.
\end{eqnarray}
The first two terms in Eq.(\ref{hamiltonian}) are the Hamiltonians of 
an STM tip (normal lead) 
 and two SC leads, accordingly:
\begin{eqnarray}\label{hamleads}
&&H_N=\sum_{k\sigma} (\varepsilon_k-eV)c^{\dag}_{k\sigma}c_{k\sigma},\\
&&H_{S}=\sum_{kj \sigma}\left\{\xi_{kj} a^{\dag}_{kj\sigma }a_{kj\sigma }-\Delta e^{i \phi_j}(a^{\dag}_{kj\uparrow }a^{\dag}_{-kj\downarrow }+H.c.)\right\}. 
\end{eqnarray}
Here,  $c_{k\sigma}$ ($c_{k\sigma}^{\dag}$) and $a_{kj\sigma}$ ($a_{kj\sigma}^{\dag}$)  are annihilation (creation) operators of 
electrons in the normal
 and $j$-th SC leads ($j=L,R$) with energies $\varepsilon_k$ and $\xi_{kj}$, correspondingly. The index $\sigma=\uparrow,\downarrow$ indicates the spin of electrons in the leads.

The Hamiltonian of the single-level vibrating CNT-QD reads as follows
\begin{eqnarray}\label{hamQD}
H_{CNT}=\sum_{\sigma}&&\varepsilon_0 d^{\dag}_{\sigma}d_{\sigma}+\frac{\hbar\omega_0}{2}(\hat p^2 + \hat x^2)
-F \hat x \sum_{\sigma}n_{\sigma}.
\end{eqnarray}
The quantum dynamics of  the electronic degree of freedom is described by the first term in Eq. (\ref{hamQD}), where $\varepsilon_0$ is the QD electron energy level, and $d_{\sigma}$, $d_{\sigma}^{\dag}$ are annihilation and creation operators of the electrons in the QD, $n_{\sigma}=d^{\dag}_{\sigma}d_{\sigma}$ \cite{footnote}. 

The second term in Eq.~(\ref{hamQD}) characterizes the CNT vibrations  with the frequency $\omega_0$, and the dimensionless operators $\hat x=\hat X/x_0$, $\hat p=x_0 \hat P/\hbar$ are canonically conjugated displacement and momentum of the CNT-QD. Here,  $x_0=\sqrt{\hbar/m\omega_0}$ is the amplitude of the zero-point oscillations of the CNT, and $m$ is the mass of the CNT. Electromechanical interaction determined by the third term in Eq.~(\ref{hamQD}), is achieved through the electrostatic interaction of the charged CNT-QD with 
the gate electrode. 
The interaction strength is $F\propto (e x_0/h) V_g\beta $ \cite{blanter},\cite{blanterbook}, where 
$h$ is the distance between the CNT and gate electrode, and 
$\beta\sim0.1$ is a geometrical factor associated with the capacitances in the system.

The last 
term in Eq.~(\ref{hamiltonian}),
\begin{eqnarray}\label{tun}
H_{tun}=\sum_{k\sigma}e^{- \hat x/\lambda}\left(t_k^n c^{\dag}_{k\sigma}d_{\sigma}+(t_k^n)^{\ast} d^{\dag}_{\sigma}c_{k\sigma}\right)\nonumber\\
+\sum_{kj\sigma}\left(t_k^s a^{\dag}_{kj\sigma}d_{\sigma}+(t_k^s)^{\ast} d^{\dag}_{\sigma}a_{kj\sigma}\right),
\end{eqnarray}
describes the tunneling processes 
between the CNT and i) 
the STM tip with deflection dependent hopping amplitude, i.e. $t_k^n \exp(-\hat x/\lambda)$, where $\lambda=l/x_0$ and $l$ is the 
tunneling length of the barrier; ii) SC leads  with the hopping amplitude  $t_k^s$.

{\it Mechanical instability.}  In order to rigorously demonstrate the phenomenon of mechanical instability in the SC hybrid junction, we analyze the dynamics 
of the CNT's flexural mode by using the reduced density matrix technique. By treating the tunneling Hamiltonian (\ref{tun}) as a perturbation and tracing out the electronic degrees of freedom in the normal and SC leads, one can get a quantum master equation for the reduced density matrix operator (in $\hbar=1$ units):
\begin{eqnarray}\label{mast}
\dot \rho =&& -i[H_{CNT},\rho]+i\Gamma_S(\phi)[d^{\dag}_{\uparrow}d^{\dag}_{\downarrow}+d_{\downarrow}d_{\uparrow},\rho]-\sum_{\sigma}\mathcal{L}[\rho].
\end{eqnarray}
Here, $\Gamma_S(\phi)=2\pi \nu_0 |t_k^s|^2\cos(\phi/2)$ is the Josephson phase dependent strength of the intra-QD electron pairing 
{induced by the coherent Andreev tunneling}, $\nu_0$ is the electron density of states in the leads, 
and $\mathcal{L}[\rho]$ is a Lindbladian operator in the high-voltage regime $eV\gg \varepsilon_0, \omega_0$ \cite{novotny},\cite{fedorets}:
\begin{eqnarray}\label{lindbladian}
\mathcal{L}[\rho]=\frac{\Gamma}{2}\left\{\begin{array}{c}\{e^{-\frac{2\hat x}{\lambda}}d_{\sigma}d^{\dag}_{\sigma},\rho\}-2e^{-\frac{\hat x}{\lambda}}d^{\dag}_{\sigma}\rho d_{\sigma}e^{-\frac{\hat x}{\lambda}},V>0,\\
\{e^{-\frac{2\hat x}{\lambda}}d_{\sigma}^{\dag}d_{\sigma},\rho\}-2e^{-\frac{\hat x}{\lambda}}d_{\sigma}\rho d_{\sigma}^{\dag}e^{-\frac{\hat x}{\lambda}},V<0,
\end{array}\right.
\end{eqnarray}
where $\Gamma=2\pi \nu_0 |t_k^n|^2$ is the QD energy level width produced by a single electron tunneling.
The quantum master equation (\ref{mast}) is justified in the {deep sub-gap regime} under the following assumptions: all {relevant} energies are smaller than the SC gap, $eV, k_BT, \varepsilon_0\ll \Delta$.


Density matrix $\rho$ acts in the finite Fock space of the two-fold degenerate single-electron level in the QD. The four possible electronic states are $|0\rangle$, $|\sigma\rangle=d^{\dag}_{\sigma}|0\rangle$ ($\sigma=\uparrow,\downarrow$), and $|2\rangle=d^{\dag}_{\uparrow}d^{\dag}_{\downarrow}|0\rangle$. 
In this representation the reduced density matrix $\rho$ contains five nonzero elements: $\rho_{00}$, $ \rho_{\uparrow\uparrow}$$=$$\rho_{\downarrow\downarrow}$$\equiv $$\rho_1$, $\rho_{22}$, $\rho_{02}$, and $\rho_{20}$. Using the normalization condition $\rho_{00}+2\rho_{1}+\rho_{22}=1$ one can eliminate the $\rho_1$ component of the density matrix from further consideration. {Therefore, the joint dynamics of the electronic and mechanical subsystems is determined by the matrix}
\begin{eqnarray}\label{dmo}
\hat{\varrho}=\frac{1}{2}\left(\begin{array}{cc} \rho_{22}-\rho_{00}&2\rho_{20} \\
    2 \rho_{02}&
                         \rho_{00}-\rho_{22}
                   \end{array}\right).\,
\end{eqnarray}
 
\begin{figure}
\includegraphics[width=80mm]{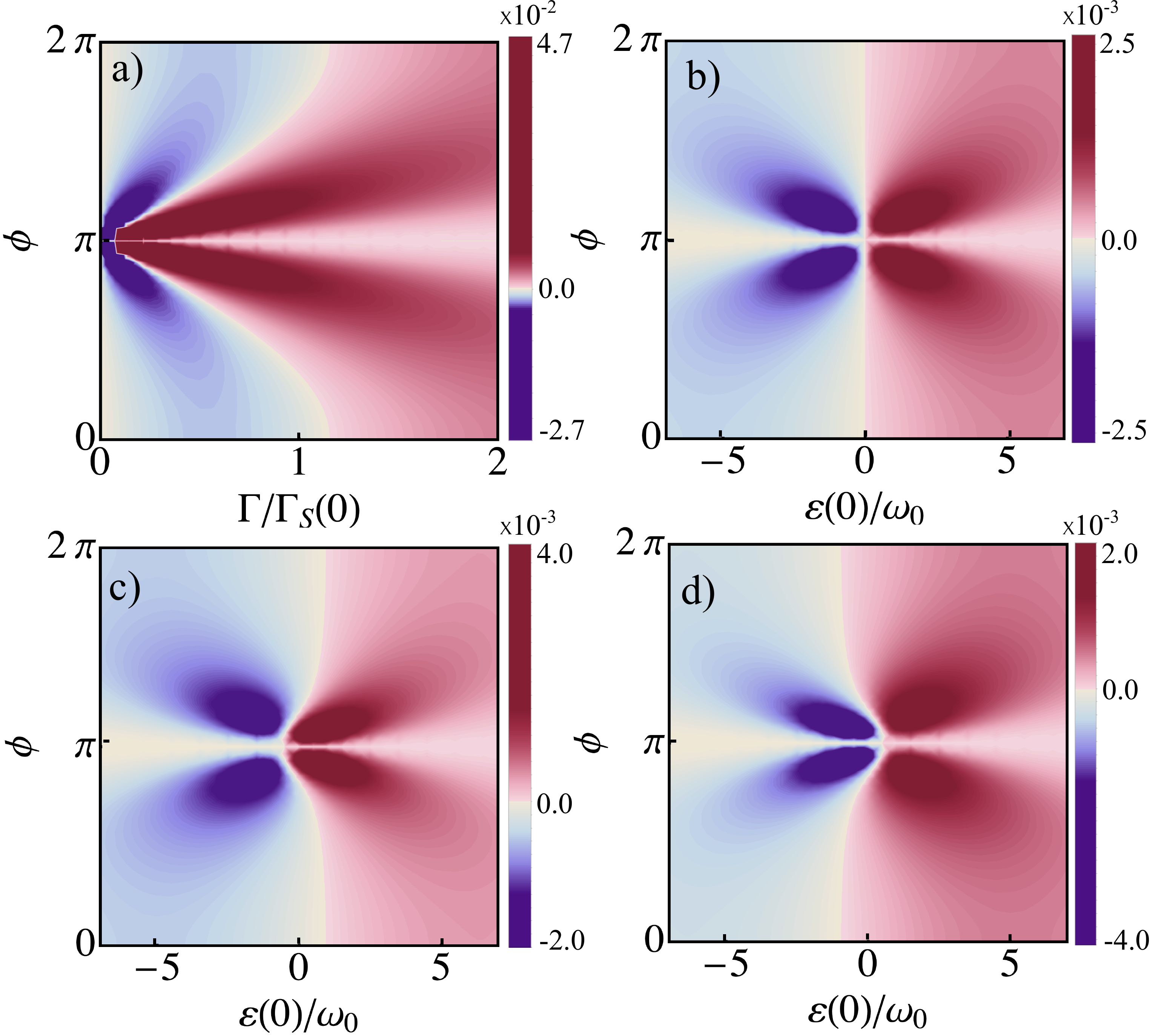}
\caption{ Phase diagrams of the mechanical instability showing pumping coefficient $\eta(0)$ as a function of the Josephson phase difference $\phi$, the QD level width $\Gamma/\Gamma_S(0)$, and the QD energy level $\varepsilon(0)/\omega_0$ for: a) $\alpha=0.2$, $\lambda^{-1}=0.05$, and $\varepsilon(0)=0$; b) $\alpha=0.2$, $\Gamma/\Gamma_S(0)=0.3$, and $\lambda^{-1}=0$; and for general case $\Gamma/\Gamma_S(0)=0.3$, $\lambda^{-1}=0.05$ when c) $\alpha=0.2$ and d) $\alpha=-0.2$.  The red and blue color schemes indicate the mechanical instability ($\eta>0$) and the damping ($\eta<0$) regimes, respectively. 
 All diagrams are calculated for the case $Q^{-1}=0$ and $\kappa=1$.
}\label{Fig2}
\end{figure}
 
 If the amplitude of the CNT displacement is larger than the amplitude of zero-point oscillation,
 one can {treat the dynamics of the CNT bending
 as a classical} with time-evolution governed by Newton's equation. { Introducing the dimensionless time units as $\omega_0t\rightarrow t$ we obtain a} closed system of the relevant equations for the CNT displacement $x$ and matrix $\hat \varrho$ Eq.~(\ref{dmo}) in the following form:
\begin{eqnarray}
&&\ddot { x}+Q^{-1}\dot x+ x=\alpha+\alpha \textrm{Tr}\{\sigma_3 \hat\varrho\},\label{newton}\\
&&{\omega_0}\dot{ \hat\varrho}=-i[\varepsilon(x)\sigma_3-\Gamma_S(\phi)\sigma_1,\hat\varrho]-\Gamma(x)\left(\hat\varrho-\frac{\kappa}{2}\sigma_3\right),\label{ME}
\end{eqnarray}
where {dimensionless parameter $\alpha=F/\omega_0$}, $\sigma_i$ ($i=1,2,3$) are the Pauli matrices, $\varepsilon(x)=\varepsilon_0-\alpha x$, $\Gamma(x)=\Gamma\exp(-2x/\lambda)$, and $\kappa=\textrm{sgn}(V)$. An environment   induced damping of the mechanical subsystem {is determined by the term $\propto Q^{-1}$, where $Q\sim10^6$ \cite{huttel1} is the quality factor.}
In the adiabatic limit, $\omega_0/\Gamma\ll 1$, we obtain 
$ \hat \varrho (t)$ from Eq.~(\ref{ME}), and the non-linear part of Eq.~(\ref{newton}) is presented in the following form:
\begin{eqnarray}\label{stationarysolution}
\textrm{Tr}\{\sigma_3 \hat \varrho(t)\} =\kappa\left(1-\frac{4\Gamma_S^2(\phi)}{D(x(t),\phi)}\right)+ \dot x(t)\eta(x(t)),
\end{eqnarray}
where $D(x,\phi)$$=$$\xi^2(x)+\Gamma^2(x)$, $\xi(x)=2\sqrt{\varepsilon^2(x)+\Gamma_S^2(\phi)}$ is the {energy difference} between two Andreev levels of the QD-SC subsystem, and a mechanical friction coefficient $\eta(x)$,
induced by interaction with the 
electronic degree of freedom, reads as
\begin{eqnarray}\label{eta}
\eta(x) =\alpha \mathcal{I}(x) \left(\lambda^{-1}C_1(x)+ \alpha \frac{\varepsilon(x)}{\Gamma^2(x)}C_2(x) \right).
\end{eqnarray}
Here, $\mathcal{I}(x)=\kappa4\Gamma(x)\Gamma_S^2(\phi)/D(x,\phi)$ is the DC flow of electrons between the STM tip and SC leads, and
\begin{eqnarray}\label{const}
C_1(x)=
\frac{6\Gamma^2(x)-2\xi^2(x)}{D^2(x,\phi)}, C_2(x)=\frac{20\Gamma^2(x)+4\xi^2(x)}{D^2(x,\phi)}.
\end{eqnarray}
The frequency of a typical CNT-based resonator is $\omega_0\sim 1~ \textrm{GHz}$, while the amplitude of zero-point fluctuations is $x_0\approx 2 ~\textrm{pm}$. Assuming $ V_g \sim 100 ~\textrm{mV}$, 
 $h\sim 10^{-7}\textrm{m}$, { and the tunneling length $l\simeq10^{-10}\textrm{m}$} we estimate dimensionless coupling constants to be $\alpha \sim 0.1$ and $\lambda^{-1}\sim 10^{-2}$.

After substituting Eq.(\ref{stationarysolution}) in Eq.(\ref{newton}), we found non-linear equation for the CNT deformation local in time.
In the limit $\alpha,\lambda^{-1}\ll 1$ a small shift of the equilibrium position (static solution) 
is obtained as
\begin{eqnarray}
x_{c}= \alpha +\kappa\alpha
\frac{4\varepsilon^2(0)+\Gamma^2}{D(0,\phi)}+O(\alpha^{2},\alpha\lambda^{-1}).
\end{eqnarray}
The stability of the  static solution is studied by linearizing Eq.~(\ref{stationarysolution}). In the limit $\Gamma\gg \omega_0$, the time evolution of the small CNT deviation from its equilibrium position $\delta x(t)=x(t)-x_c$  is given by \cite{footnote2}
\begin{eqnarray}\label{Newton3}
\delta\ddot{ x} + \left( Q^{-1} -\eta (0)\right) \delta\dot{ x} + \delta x =0.
\end{eqnarray}
The static solution $x_c$ of the system at $\eta(0)>Q^{-1}$ { becomes {\it unstable}} with respect to the generation of mechanical oscillation with amplitude exponentially increasing in time. Development of instability results in the appearance of self-sustained mechanical oscillations, governed by the nonlinearity of r.h.s. Eq.~(\ref{newton}).

{Next, we analyze the influence of various parameters on the coefficient $\eta(0)$ which we call a pumping coefficient in what follows. First, we note that $\eta(0)$ linearly increases with the electromechanical coupling $\alpha$ and the DC flow $\propto \mathcal {I}(0)$. Moreover, the pumping coefficient $\eta(0)$ changes a sign depending on the direction of the electronic flow, i.e. the sign of $eV$.}  
At $|eV|\gg 2\varepsilon_0$, bias voltage affects the phenomenon under consideration solely by this means. Below we analyze the case of $eV>0$ only.

The various dependencies of the pumping coefficient $\eta(0)$ on the parameters $\phi$, $\Gamma/\Gamma_S(0)$ and $\varepsilon(0)$ obtained from Eqs.~(\ref{eta}) and (\ref{const}) are shown in  Fig.~\ref{Fig2} (red color scheme indicates $\eta(0)>0$, while blue scheme -- $\eta(0)<0$).   
In the case $\varepsilon(0)$$=$$\varepsilon_0$$=$$0$, the pumping coefficient $\eta(0)\propto\kappa\alpha/\lambda$ is determined by the ratio between $\Gamma$ and $\Gamma_S(\phi)$, {since only the first term in Eq.~(\ref{eta}) contributes. The pumping coefficient changes its sign when $\Gamma=\sqrt{4/3}\Gamma_S(\phi)$, see Fig.~\ref{Fig2}(a).}
If the dependence of the electron hopping on the amplitude of the CNT oscillations is negligible, i.e. $\lambda^{-1}=0$, the pumping coefficient $\eta(0)\propto \kappa\alpha^2 \varepsilon(0)$ is determined by the sign of $\varepsilon(0)$. Such behavior is illustrated in Fig.~\ref{Fig2}(b). General case, when both terms in Eq.~(\ref{const}) contribute into the pumping coefficient Eq.~(\ref{eta}), is shown in Fig.~\ref{Fig2}(c) and (d) for positive ($\alpha>0$) and negative ($\alpha<0$) electrostatic interaction, respectively. 

{The origin of the pumping processes, and corresponding mechanical instability can be qualitatively explained as follows:
 since two electronic states $\vert 0\rangle$ and $\vert 2 \rangle$ in the QD are not the eigenstates of the QD-SC subsystem, the quantum Rabi oscillations emerge with a frequency proportional to the {energy difference} between Andreev levels $\xi(x,\phi)$. These Rabi oscillations occur in the form of periodic in time single-Cooper pair transfer between SC leads and the QD. However, an incoherent single electron tunneling from the STM tip to the QD can interrupt the coherent oscillations as well as resume them. 
 
 As this takes place, the averaged charge in the QD is governed by the interplay between two processes: i) a coherent Rabi oscillations and ii) an incoherent single electron tunneling. Both processes and their main characteristics, $\Gamma(x)$ and $\xi(x)$, are controlled by the CNT displacement and vary in time if  $\delta\dot x(t)\neq 0$. Such variations give rise to a correction of the average charge in the QD, that is proportional to the velocity of the QD, thereby generating effective friction force. {We note that the amplitude of the effective friction force is determined by two terms (see Eq. (\ref{eta})), where the first term is induced by the time variation of the hopping amplitude of single electron tunneling $\dot\Gamma(x(t))\propto \lambda^{-1}\dot x$, while the second term is generated by the time variation of the Rabi frequency $\dot\xi(x(t))\propto \alpha \varepsilon(0)\dot x$.


{\it DC electric current.}
 The self-sustained oscillations 
affect the DC current flow between the STM tip and SC leads. This phenomenon allows one to verify the mechanical instability through the electric current measurement. 
\begin{figure}
\includegraphics[width=80mm]{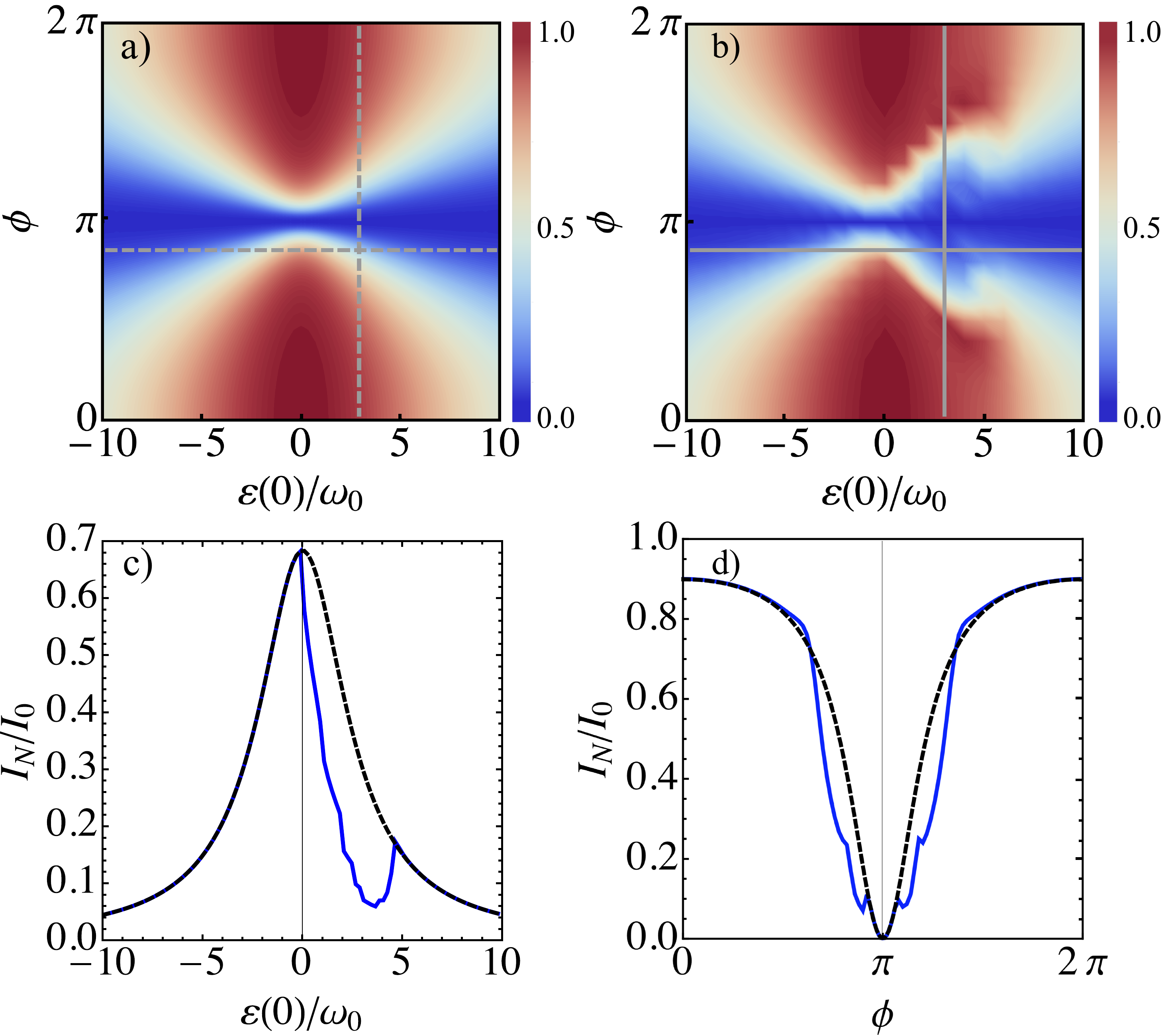}
\caption{  DC electric current $I_N/I_0$ normalized to the maximum of static current $I_0=e\Gamma$ as a function of the Josephson phase difference $\phi$ and the QD energy level $\varepsilon(0)/\omega_0$ at $\Gamma/\Gamma_S(0)=0.3$ for the cases: a) $\alpha=0$, and b) $\alpha=0.2$. {Dashed and solid grey lines indicate projections of the DC current at fixed $\phi=2.7$ and fixed $\varepsilon(0)/\omega_0=3$, respectively.} These projections are presented in panels c) and d), where the charge current ($I_N(0)=e\mathcal{I}(0)$) at $\alpha=0$ is shown by black dashed lines, and the DC current at $\alpha=0.2$ is shown by the blue (solid) lines.  Current in the pumping regime is calculated numerically from Eqs.~(\ref{curr}),(\ref{newton}),(\ref{ME}) by averaging over the period of mechanical vibrations. All figures are obtained for $Q=10^6$, $\kappa=1$, and $\lambda^{-1}=0.05$.}
\label{Fig3}
\end{figure}

The expression for the DC current is given by
\begin{eqnarray}\label{curr}
I_N(x(t))=e\Gamma\left(x(t)\right) \left(\kappa-\textrm{Tr}\{ \sigma_3 \hat \varrho(t)\}\right).
\end{eqnarray}
If the pumping coefficient $\eta(0)<Q^{-1}$,  the  mechanical oscillations of the CNT are damped, and  the DC electric current is expressed as $I_N(0)=e\mathcal{I}(0)$. This expression coinsides with the DC current obtained in the absence of electromechanical interaction. Such dependence is shown in Fig.~\ref{Fig3}(a). The DC current strongly depends on the Josephson phase difference $\phi$ and the QD energy level $\varepsilon(0)$. 
The current reaches its maximum at $\varepsilon(0)=0$ and vanishes at $\phi=\pi$.
Besides, $I_N(0)$ is proportional to $\propto \Gamma \Gamma_S^2$, revealing Andreev tunneling \cite{andreevtunneling} since only two electrons (the Cooper pair) can tunnel from the QD to the SC leads. 

In the regime of mechanical instability $\eta(0)>Q^{-1}$, the static solution becomes  unstable and CNT vibrations develop into pronounced self-sustained oscillations of finite amplitude.
As a result, the current exhibits periodic oscillations with the frequency $\omega_0$. The averaged over the period of mechanical oscillations DC current 
 is obtained numerically and the result is presented in Fig.~\ref{Fig3}(b). The projections of $I_N$ at fixed $\phi$
  and 
  $\varepsilon(0)$ are presented in Fig.~\ref{Fig3} (c) and (d). 
As one can see in Fig.~\ref{Fig3}, 
pronounced self-sustained oscillations of the CNT-QD suppress the charge current in the region of parameters obeyed $\eta(0)>Q^{-1}$ condition. The strength of this current suppression depends on the amplitude of the CNT self-oscillations and correspondingly on the pumping strength $\eta(0)$. 

\textit{Conclusions.} { We predict the phenomenon of mechanical instability and corresponding self-sustained oscillations in a hybrid nanoelectromechanical device consisting of a carbon nanotube suspended between two SC leads and placed near a voltage-biased normal STM tip. Such effect is based on a peculiar interplay of the coherent quantum-mechanical Rabi oscillations induced by the Andreev tunneling between the CNT and SC leads, and an incoherent single electron tunneling between the STM tip and CNT. We obtain that the observed mechanical instability {and self-sustained oscillations of finite amplitude are} determined by two parameters:  
the relative position of the single-electron energy level, and
the Josephson phase difference between the SC leads. 
Numerical analysis demonstrates that
the predicted mechanical instability develops
into pronounced self-sustained bending oscillations of the CNT
resonator which, in its turn, result in a suppression of the
DC electric current flowing between the STM tip and SC leads. This effect allows one to detect the predicted mechanical
instability through the DC current measurement.
A SQUID sensitivity
to an external magnetic field can be achieved by using proposed nanomechanical
Andreev device through the control of the Josephson phase difference by a
magnetic flux.

{\it Acknowledgement.} 
This work was supported by the Institute for Basic Science in Korea (IBS-R024-D1). LYG and RIS thank the IBS Center for Theoretical Physics of Complex Systems for their hospitality.
 M.V.F. acknowledges the partial financial support of the Ministry of Education and Science of the Russian Federation in the framework of Increase Competitiveness Program of NUST "MISIS" $K2-2020-001$.

\vspace*{3mm}



\begin{thebibliography}{99}
\bibitem{footnote}
We neglect the inter-dot electron interaction $U n_{\uparrow}n_{\downarrow}$ in case when $U<eV$, since it will give only qualitative correction to the phenomenon under consideration. More precisely, it will renormalize QD energy level, $\varepsilon_0\rightarrow \varepsilon_0+U/2$.
\bibitem{footnote2}
In Eq.~(\ref{Newton3}) we ignore small renormalization of the frequency of mechanical oscillations.
\end{thebibliography}
\end{document}